\documentclass[preprint,tightenlines,aps,floats,nofootinbib]{revtex4}

\usepackage{epsf}
\usepackage{epsfig}
\usepackage{graphicx}
\usepackage[dvips]{color}

\newcommand{\notE}{\ \hbox{{$E$}\kern-.60em\hbox{/}}}
\newcommand{\notp}{\ \hbox{{$p$}\kern-.43em\hbox{/}}}
\def\D0{\mbox{D\O}}

\newcommand{\Del}{\Delta}
\newcommand{\eps}{\epsilon}

\includegraphics{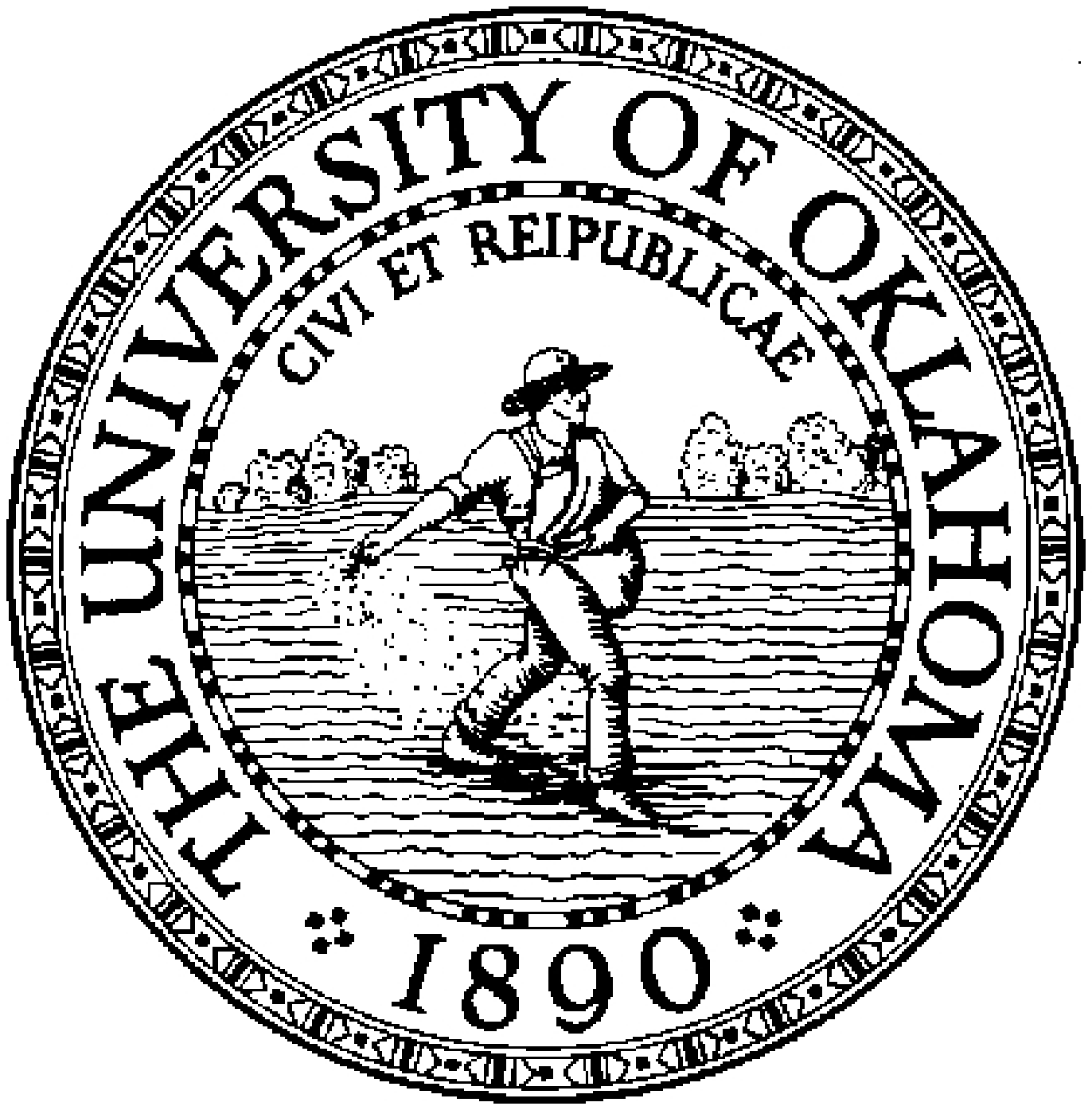}

\preprint{\font\fortssbx=cmssbx10 scaled \magstep2
\hbox to \hsize{
\hskip1.2in 
\hbox{\fortssbx The University of Oklahoma}
\hskip0.2in $\vcenter{
                      \hbox{\bf arXiv:0908.1156 [hep-ph]}
                      \hbox{October 2009}}$ }
}

\begin{document}
 
\title{\vspace*{0.7in}
Discovering the Higgs Bosons of Minimal Supersymmetry \\ 
with Bottom Quarks}

\author{
Chung Kao\thanks{Email address: kao@physics.ou.edu}, 
Shankar Sachithanandam,
Joshua Sayre\thanks{Email address: sayre@physics.ou.edu}, and 
Yili Wang}

\affiliation{
Homer L. Dodge Department of Physics and Astronomy \\
University of Oklahoma \\ 
Norman, Oklahoma 73019, USA 
\vspace*{.4in}}


\thispagestyle{empty}

\begin{abstract}

We investigate the prospects for the discovery of a neutral Higgs boson 
produced with one bottom quark followed by Higgs decay into 
a pair of bottom quarks at the CERN Large Hadron Collider (LHC) 
and the Fermilab Tevatron Collider.  
We work within the framework of the minimal supersymmetric standard model.
The dominant physics background is calculated with realistic
acceptance cuts and efficiencies including the production of 
$bb\bar{b}$, $\bar{b}b\bar{b}$, 
$jb\bar{b}$ ($j = g, q, \bar{q}$; $q = u, d, s, c$), 
$t\bar{t} \to b\bar{b}jj\ell\nu$, and $t\bar{t} \to b\bar{b}jjjj$.
Promising results are found for the CP-odd pseudoscalar ($A^0$) 
and the heavier CP-even scalar ($H^0$) Higgs bosons with masses 
up to 800 GeV for the LHC with an integrated luminosity ($L$)
of 30 fb$^{-1}$ and up to 1 TeV for $L =$ 300 fb$^{-1}$.

\end{abstract}


\maketitle

\newpage

\color{black}

\section{Introduction}

The Fermilab Tevatron Run II has been taking data since March 2001, 
and the CERN Large Hadron Collider (LHC) is planned to start running 
in Autumn 2009.
One of the most important experimental goals of the Tevatron Run II and 
the LHC is the search for the mechanism of electroweak symmetry
breaking---to discover the Higgs bosons or to prove their non-existence.

In the Standard Model, only one Higgs doublet is required to generate 
mass for both gauge bosons and elementary fermions, and the Higgs boson
is the only particle remaining to be discovered in high energy experiments.
In the minimal supersymmetric standard model (MSSM) \cite{MSSM}, 
the Higgs sector has Yukawa interactions with two doublets, 
$\phi_1$ and $\phi_2$, whose neutral components couple to fermions with
weak isospin 
$t_3 = -1/2$ and $t_3 = +1/2$ respectively \cite{Guide}. 
After spontaneous symmetry breaking, there remain five physical Higgs bosons:
a pair of singly charged Higgs bosons $H^{\pm}$,
two neutral CP-even scalars $H^0$ (heavier) and $h^0$ (lighter),
and a neutral CP-odd pseudoscalar $A^0$.
The Higgs potential is constrained by supersymmetry 
such that all tree-level Higgs boson masses and couplings 
are determined by just two independent parameters,  
commonly chosen to be the mass of the CP-odd pseudoscalar ($M_A$) 
and the ratio of vacuum expectation values of neutral Higgs fields 
($\tan\beta \equiv v_2/v_1$). 

At the LHC, 
gluon fusion ($gg \to \phi^0;$ $ \phi^0 = h^0, H^0, A^0$) is the major 
source of neutral Higgs bosons in the MSSM for $\tan\beta$ less than 5.
For $\tan\beta > 7$, neutral Higgs bosons are dominantly 
produced from bottom quark fusion $b\bar{b} \to \phi^0$ 
\cite{Dicus1,Dicus2,Balazs,Maltoni,Harlander}. 
Since the Yukawa couplings of $\phi^0b\bar{b}$ are enhanced by $1/\cos\beta$,
the production rate of neutral Higgs bosons associated with bottom
quarks, especially that of the $A^0$ or the $H^0$, 
is enhanced at large $\tan\beta$.

For a Higgs boson produced along with a single bottom quark at high transverse 
momentum ($p_T$), the leading-order subprocess is 
$bg \to b \phi^0$ \cite{Choudhury,Huang,Scott2002,Cao,Dawson:2007ur}. 
If two high $p_T$ bottom quarks are required in association with 
a Higgs boson, the leading order subprocess should be
$gg \to b\bar{b}\phi$ \cite{Dicus1,hbbmm,Plumper,Dittmaier,Dawson:2003kb}. 
In 2002, it was suggested that the search at the LHC for 
a Higgs boson produced along with a single bottom quark with large $p_T$ 
should be more promising than the production of a Higgs boson associated with 
two high $p_T$ bottom quarks \cite{Scott2002}. 
This has already been shown to be the case for the $\mu^+ \mu^-$ 
decay mode of the Higgs bosons\cite{hbmm}. 

For large $\tan\beta$, the $\tau^+\tau^-$ decay mode \cite{Kunszt,Richter-Was}
can be a promising discovery channel for the $A^0$ and the $H^0$ 
in the MSSM. Recently, the discovery channel 
$b\phi^0 \to b \tau^+\tau^-$ has been demonstrated 
to offer great promise at the LHC to search for the $A^0$ and the
$H^0$ up to $M_A = 1$ TeV \cite{hbll}. 


The Higgs decay into $b\bar{b}$ has the largest branching fraction 
for large values of $\tan\beta$. However, the inclusive channel of 
$pp \to \phi^0 \to b\bar{b} +X$ is very challenging at the LHC 
owing to the extremely large QCD background. 
Previous theoretical studies have focused on the associated production of 
$b\bar{b}\phi^0 \to b\bar{b}b\bar{b}$ \cite{Vega,Yuan,Carena:1998gk}.
Realistic simulations by the ATLAS and the CMS collaborations 
with parton showering lead to pessimistic results 
\cite{ATLAS,ATLAS-thesis,CMS},  
because the trigger for the $4b$ final state requires high $p_T$
bottom quarks for $pp \to b\bar{b}\phi^0 \to b\bar{b}b\bar{b} +X$. 
The requirement of four high $p_T$ $b$-quarks removes most of the Higgs events.
Moreover, integrating over the fourth $b$-quark to study a $3b$ signal
requires a careful inclusion of higher order corrections in the four-flavor
scheme. These potentially large leading-log corrections are absorbed into the
$b$-quark PDFs in the five flavor scheme which we employ. 

 
In this article, we present the prospects for discovering the MSSM neutral 
Higgs bosons produced with a single high $p_T$ bottom quark ($b$ or
$\bar{b}$) followed by Higgs decay into a pair of bottom quarks .
We calculate the Higgs signal and the dominant Standard Model (SM) 
backgrounds with exact matrix elements as well as realistic cuts 
and efficiencies. Furthermore, we present promising 
$5\sigma$ discovery contours at the LHC in the $(M_A,\tan\beta)$ plane.
Section II shows the production cross sections and branching fractions
for the Higgs signal.
The SM physics background is discussed in Section III.
Sections IV and V present the discovery potential at the LHC 
and the Fermilab Tevatron Run II.
Optimistic conclusions are drawn in Section VI.

\section{The production cross sections and branching fractions}

At the LHC or the Tevatron Run II, the production cross section of 
$bg \to b\phi^0 \to b b\bar{b}$, where $ \phi^0 = H^0, h^0, A^0$, 
is evaluated with 
the parton distribution functions of CTEQ6L1 \cite{CTEQ6} 
and the factorization scale $\mu_F = M_H/4$ \cite{Scott2002}. 
In this article, $b$ represents a bottom quark ($b$) or 
a bottom anti-quark ($\bar{b}$) unless it is explicitly specified.
The bottom quark mass in the $\phi^0b\bar{b}$ Yukawa coupling 
is chosen to be the next-to-leading-order (NLO) running mass 
$m_b(\mu_R)$ \cite{bmass}, 
which is calculated with $m_b({\rm pole}) = 4.7$ GeV 
and the NLO evolution of the strong coupling \cite{alphas}. 
We have also taken the renormalization scale to be $M_H/4$.  
This choice of scale effectively reproduces the effects of next-to-leading 
order (NLO) corrections \cite{Scott2002}. Therefore, we take the $K$ factor to be one 
for the Higgs signal. 

At the LHC, we calculate the Higgs cross section 
$\sigma(pp \to b\phi^0 \to bb\bar{b} +X)$ 
with a Breit-Wigner resonance
via $bg \to b\phi^0 \to bb\bar{b}$.
In addition, we check the cross section 
with the narrow width approximation 
\begin{eqnarray*}
 \sigma(pp \to b\phi^0 \to bb\bar{b} +X) 
= \sigma(pp \to b\phi^0 +X) \times B(\phi^0 \to b\bar{b})
\end{eqnarray*}
where $B(\phi^0 \to b\bar{b})$ is the branching fraction of 
a Higgs boson decay into $b\bar{b}$.

 
\begin{figure}[htb]
\centering\leavevmode
\epsfxsize=3.6in
\epsfbox{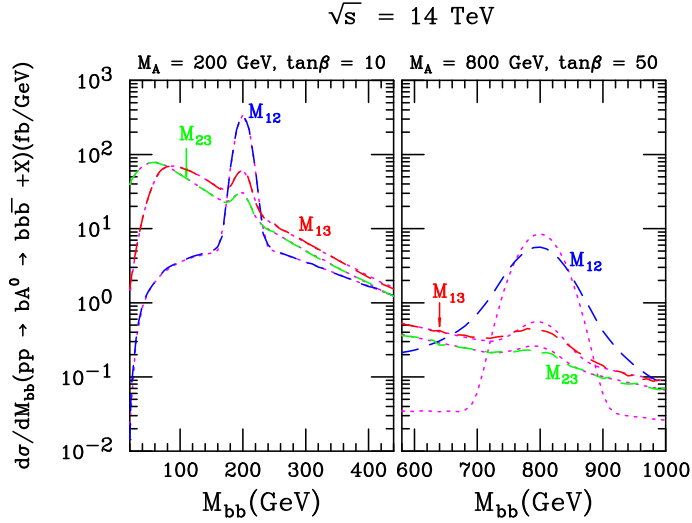}
\caption[]{
The invariant-mass distribution of $b\bar{b}$ and $bb$ pairs 
$d\sigma/dM_{bb}(pp \to b b\bar{b} +X)$, 
for the Higgs signal from $bg \to bA^0$ with $M_A = 200$ GeV 
and $\tan\beta = 10$ 
as well as $M_A = 800$ GeV for $\tan\beta = 10$ and $\tan\beta = 50$. 
We calculate the Higgs signal with a Breit-Wigner resonance (dash) and 
in the narrow width approximation (dot), applying minimal cuts of
 $p_T > 10$ GeV and $|\eta| < 10$.
\label{fig:hbbb-mass}
}
\end{figure}

Figure 1 shows the invariant mass distribution ($M_{ij}$, i,j = 1,2,3) 
of the $b_i b_j$ or $b_i \bar{b}_j$ pairs 
for the Higgs signal $pp \to bA^0 \to bb\bar{b} +X$ via $bg \to bA^0$.
The bottom quarks are ordered according to their transverse momenta, 
$p_T(b_1) \ge p_T(b_2) \ge p_T(b_3)$.
We note that with energy-momentum smearing, the cross section in
the narrow width approximation (NWA)
agrees very well with that evaluated via a Breit-Wigner resonance
(BWR) for most parameters that we have chosen.
Based on the ATLAS \cite{ATLAS}  
 specifications, 
we model these effects by Gaussian smearing of momenta:
\begin{eqnarray}
\frac{\Delta E}{E} = \frac{0.60}{\sqrt{E}} \oplus 0.03
\end{eqnarray} 
for jets at the LHC, with individual terms added in quadrature. For the Tevatron we use 
\begin{eqnarray}
\frac{\Delta E}{E} = \frac{0.50}{\sqrt{E}} \oplus 0.03
\end{eqnarray} based on  CDF parameters \cite{CDF-Abulencia:2005aj}. 
For $M_A = 800$ GeV and $\tan\beta = 50$, the cross sections 
are in agreement within $10\%$. For large values of $M_A$, the increased 
width of the Higgs may lead to a reduced signal due to cuts on the dijet 
invariant-mass acceptance window. This effect is less well-modeled in the NWA
than with BWR,
although the total cross-sections remain in good agreement.

\section{The Physics Background}

The final state of $bb\bar{b}$ has dominant physics backgrounds coming from 
(a) $bg \to b b\bar{b}$, 
(b) $cg \to c b\bar{b}$, 
(c) $qg \to q b\bar{b}$ with $q = u,d, s$,
(d) $q\bar{q} \to g b\bar{b}$ with $q = u, d, s, c$, and 
(e) $gg,q\bar{q} \to t\bar{t} \to b\bar{b} jj \ell \nu$, or
    $gg,q\bar{q} \to t\bar{t} \to b\bar{b} jj jj$.
We have computed the cross section of the Higgs signal and physics 
background utilizing MadGraph \cite{Madgraph,Madgraph2} and 
HELAS \cite{Helas} to generate matrix elements. 
To reduce the physics background while keeping most of the signal
events, we require that in each event there are three jets 
(at least two $b$-jets) which satisfy the following requirements: 
\begin{itemize}
\item[(a)] we consider two sets of cuts for an integrated luminosity ($L$) of 
30 fb$^{-1}$ (low luminosity, LL): 
(i) $p_T(j_1) > 50$ GeV, $p_T(j_2) > 30$ GeV and $p_T(j_3) > 20$ GeV 
(low $p_T$ cuts), or
(ii) $p_T(j_1,j_2,j_3) > 70$ GeV (CMS 3-jet trigger) \cite{CMS} 
as well as the pseudorapidity, $|\eta| < 2.5$ for all jets,
where $p_T(j_1) > p_T(j_2) > p_T(j_3)$, or
\item[(b)]for $L = 300$ fb$^{-1}$ (high luminosity, HL) we check two sets of cuts : 
(i) $p_T(j_1,j_2,j_3) > 75$ GeV (ATLAS 3-jet trigger) \cite{ATLAS-thesis} 
or 
(ii) $p_T(j_1,j_2,j_3) > 150$ GeV (ATLAS 3-jet trigger for high luminosity) 
\cite{ATLAS-thesis} as well as $|\eta| < 2.5$ for all jets,
\item[(c)] there is at least one pair of bottom quarks 
in the Higgs mass window such that \\
$|M_{bb} - M_\phi| < \Del M_{bb}$, 
where $\Del M_{bb} = {\rm MAX}(22 \,{\rm GeV}, \sigma_M)$, choosing 
$\sigma_M = 0.10 \times M_\phi$ or $0.15 \times M_\phi$ 
for $L = 30$ fb$^{-1}$ and 
$\sigma_M = 0.15 \times M_\phi$ or $0.20 \times M_\phi$ 
for $L = 300$ fb$^{-1}$,
\item[(d)] all three jets are separated with $\Delta R =
  \sqrt{\Delta\phi^2+\Delta \eta^2} > 0.7$ (where $\phi$ is the angle
between two jets in the transverse plane), 
\item[(e)] the missing transverse energy ($\notE_T$) should be less
  than 40 GeV.
\end{itemize}

In addition, we veto events with more than three jets passing the cuts
$p_T(j) > 15$ GeV and $|\eta| < 2.5$.
We take the $b-$tagging efficiency to be $\eps_b = 0.6$ (LL) or 
$\eps_b = 0.5$ (HL),  
the probability that charm quark may be misidentified is $\eps_c = 0.15$, 
and the probability that a light quark or a gluon may be misidentified as 
a bottom quark is $\eps_j = 0.01$.
For the backgrounds arising from  $bb\bar{b}$ and $jb\bar{b}$ \cite{Vega} 
as well as those from $t\bar{t}$ \cite{ttbar}, we assume a $K$ factor
of $2$ when computing the significance as discussed below. 
In practice we find that the $t\bar{t}$ backgrounds are negligible
after cuts, although we include them for completeness.

\bigskip

In Figure 2, we present the transverse momentum distribution
($d\sigma/d{p_T}$) of the bottom quarks ($b$ or $\bar{b}$),
for the Higgs signal 
$pp \to bA^0 \to bb\bar{b} +X$.
Also shown is the $p_T$ distribution for bottom quarks from the SM background 
$bg \to bb\bar{b}$.
We have required $p_T(b) > 10$ GeV and $|\eta_b| < 2.5$ in this figure.

 
\begin{figure}[htb]
\centering\leavevmode
\epsfxsize=3.6in
\epsfbox{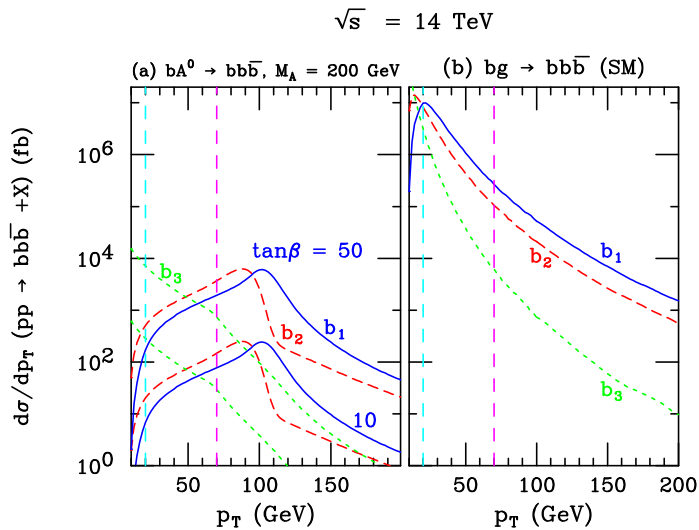}
\caption[]{
The transverse-momentum distribution for 
(a) the Higgs signal from $bg \to bA^0$ with $M_A = 200$ GeV and 
 $\tan\beta = 10, 50$ 
as well as for (b) the physics background from $bg \to b b\bar{b}$.
We require $p_T(b) > 10$ GeV and $|\eta_b| < 2.5$ in this figure.
 The vertical, dashed lines illustrate cuts at $20$ GeV and $70$ GeV.
\label{fig:hbbb-momentum}
}
\end{figure}

\section{The Discovery Potential at the LHC}

To study the discovery potential of 
$pp \to b \phi^0 \to b b\bar{b} +X$ $ ( \phi^0 = H^0, h^0, A^0)$ at the LHC, 
we calculate the Higgs signal as well as the SM physics background 
in the mass window of
$M_\phi \pm \Delta M_{bb}$ where 
$\Delta M_{bb} = {\rm MAX}(22 \, {\rm GeV},\, 0.10 \times M_\phi)$, or 
$\Delta M_{bb} = {\rm MAX}(22 \, {\rm GeV},\, 0.15 \times M_\phi)$ 
for an integrated luminosity of 30 fb$^{-1}$.

 
\begin{figure}[htb]
\centering\leavevmode
\epsfxsize=3.6in
\epsfbox{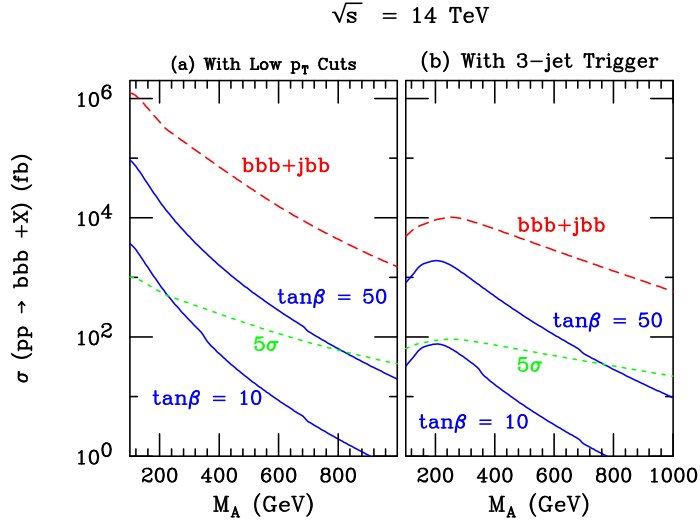}
\caption[]{
The signal cross section of $bg \to bA^0$ at the LHC for an integrated
luminosity $L = 30$ fb$^{-1}$, as a function of $M_A$, for
$m_{\tilde{q}} = m_{\tilde{g}} = \mu = 1$ TeV, $\tan\beta = 10$ 
and $\tan\beta = 50$. 
Also shown are the background cross sections in the mass window of 
$M_A \pm 0.10 \times M_A$ as discussed in the text for the SM contributions. 
We have applied acceptance cuts and efficiencies of 
tagging and mistagging. 
\label{fig:hbbb-sigma}
}
\end{figure}

In Figure 3 we show the cross section of 
$\sigma(pp \to b A^0 \to b b\bar{b} +X)$, 
for $\tan\beta = 10$ and 50, with a common mass for scalar quarks, scalar 
leptons and the gluino $m_{\tilde{f}} = m_{\tilde{g}} = \mu = 1$ TeV.
We also present the background cross sections with no $K$ factor 
in the mass window of 
$M_A \pm \Delta M_{bb}$ for the dominant SM processes
$pp \to b b\bar{b} +X$ and $pp \to j b\bar{b} +X, j = q, \bar{q}, g$, 
with (a) low $p_T$ cuts and (b) CMS 3-jet trigger.
The cuts and tagging efficiencies are included 
with $\Delta M_{bb} = 0.10\times M_A$. 
In addition, we present the $5\sigma$ cross section 
for $L = 30$ fb$^{-1}$.
The cross section of the Higgs signal with $\tan\beta \simeq 50$ 
can be larger than the $5\sigma$ cross section for $M_A \alt 800$ 
after acceptance cuts.  
Requiring higher transverse momenta ($p_T > 70$ GeV) greatly reduces 
the background and the Higgs signal for $M_A < 200$ GeV.

We define the signal to be observable 
if the lower limit on the signal plus background is larger than 
the corresponding upper limit on the background \cite{HGG,Brown}, namely,
\begin{eqnarray}
L (\sigma_s+\sigma_b) - N\sqrt{ L(\sigma_s+\sigma_b) } > 
L \sigma_b +N \sqrt{ L\sigma_b }
\end{eqnarray}
which corresponds to
\begin{eqnarray}
\sigma_s > \frac{N^2}{L} \left[ 1+2\sqrt{L\sigma_b}/N \right]
\end{eqnarray}
Here $L$ is the integrated luminosity, 
$\sigma_s$ is the cross section of the Higgs signal, 
and $\sigma_b$ is the background cross section.  
Both cross sections are taken to be 
within a bin of width $\pm\Delta M_{bb}$ centered at $M_\phi$. 
In this convention, $N = 2.5$  corresponds to a 5$\sigma$ signal.
We take the integrated luminosity $L$ to be 30 fb$^{-1}$ 
and 300 fb$^{-1}$ \cite{ATLAS}. 

For $\tan\beta \agt 10$, 
$M_A$ and $M_H$ are almost degenerate when $M_A \agt$ 125 GeV, 
while $M_A$ and $m_h$ are very close 
to each other for $M_A \alt$ 125 GeV in the MSSM \cite{Higgs-Mass}. 
Therefore, when computing the discovery reach, we add the cross sections 
of the $A^0$ and the $h^0$ for $M_A < 125$ GeV 
and those of the $A^0$ and the $H^0$ for $M_A \ge 125$ GeV 
\cite{Nikita,CMS,ATLAS}.

Figure 4 shows the 5$\sigma$ discovery contours for the MSSM Higgs bosons 
where the discovery region is the part of the parameter space above the 
contour. We have chosen 
$M_{\rm SUSY} = m_{\tilde{q}} = m_{\tilde{g}} = m_{\tilde{\ell}} 
= \mu = 1$ TeV.
If $M_{\rm SUSY}$ is smaller, the discovery region of $A^0,H^0 \to b\bar{b}$ 
will be slightly reduced for $M_A \agt 250$ GeV,
because  the Higgs bosons can decay into supersymmetric (SUSY)
particles \cite{HZ2Z2} 
and the branching fraction of $\phi^0 \to b\bar{b}$ is suppressed.
For $M_A \alt 125$ GeV, the discovery region of $H^0 \to b\bar{b}$ 
is slightly enlarged for a smaller $M_{\rm SUSY}$, 
but the observable region of $h^0 \to b\bar{b}$ is slightly reduced 
because the lighter top squarks make the $H^0$ and the $h^0$ lighter; 
also the $H^0 b\bar{b}$ coupling is enhanced 
while the $h^0 b\bar{b}$ coupling is reduced~\cite{Nikita}.

 
\begin{figure}[htb]
\centering\leavevmode
\epsfxsize=3.6in
\epsfbox{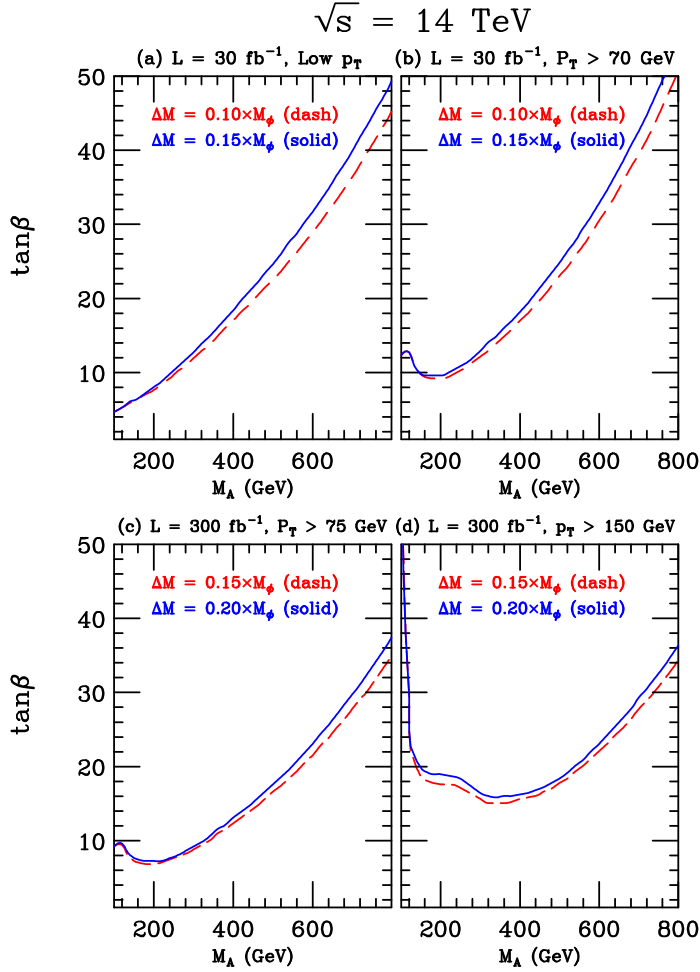}
\caption[]{
The $5\sigma$ discovery contours at the LHC with $\sqrt{s} =$ 14 TeV
for 
(a) $L = 30$ fb$^{-1}$ and low $p_T$ cuts, 
(b) $L = 30$ fb$^{-1}$ and $p_T > 70$ GeV,
(c) $L = 300$ fb$^{-1}$ and $p_T > 75$ GeV,
(d) $L = 300$ fb$^{-1}$ and $p_T > 150$ GeV,
in the $M_A$ versus $\tan\beta$ plane.  
The signal includes $\phi^0 = A^0$ and $h^0$ for $M_A < 125$ GeV, 
and $\phi^0 = A^0$ and $H^0$ for $M_A \ge 125$ GeV. 
The discovery region is the part of the parameter space above the contours.}
\label{fig:hbbb-contour1}
\end{figure}

In addition, we have studied the effect of an invariant mass cut,  
using only the two jets with highest $p_T$ as the candidate pair. 
Table I presents the cross section corresponding to two schemes:
 (a) requiring $|M_{12} -M_\phi| < \Del M_{bb}$, and
 (b) requiring $|M_{ij} -M_\phi| < \Del M_{bb}; i,j = 1,2,3$.
We find that for $M_A \agt 400$, it is more advantageous to apply 
an invariant mass cut only on the two leading $b$ jets. For lower masses
using any pair of the three leading jets leads to higher significance. We also
show the ratio of signal to background in this figure. We have chosen a set of
 cuts, $p_T(j_1,j_2,j_3) > 100,80,70$ GeV, which tends to maximize this ratio. Less 
stringent cuts can improve the nominal statistical significance in the low mass 
regions, but for high masses and low $\tan  \beta$ the small signal to background
 ratio would require excellent understanding of backgrounds and systematic errors.


\begin{table}
\label{Invariant-Mass}
\caption[]{
Cross sections in fb for the Higgs signal and physics background 
for two choices of cuts on the invariant mass of $b\overline{b}$: 
(a)  two leading jets ($M_{12}$) versus
(b) any two jets ($M_{ij}$) used to reconstruct the Higgs invariant
mass. Significances are computed with $L = 30 \; \text{fb}^{-1}$.
}
\begin{tabular}{|c |c|c|c|c|c|}
\hline
\bf{Mass}(GeV) & &\bf{Signal} & \bf{Background} & \bf{Significance} ($N_{SS} = N_S/\sqrt{N_B + N_S}$) & $N_S/N_B$\\
\hline
\multicolumn{6}{c}{$\tan  \beta = 10$} \\ \hline

$200$ & $M_{12}$  & $ 44.2$&$3960$ & $3.82$ &$1.12\times 10^{-2}$\\
&$M_{ij}$  & $126$ & $14500$ & $5.72$&$8.70\times 10^{-3}$ \\
\hline

$400$&$M_{12}$  & $23.5$& $6680$ & $1.57$&$3.52\times 10^{-3}$ \\
&$M_{ij}$  &$32.2$& $11900 $  & $1.61 $ &$2.70\times 10^{-3}$\\
\hline

$800$&$M_{12}$  & $1.42$& $1400$ & $0.208$ &$1.02\times 10^{-3}$\\
&$M_{ij}$  & $1.61 $& $2380$ & $0.181 $ &$6.76\times 10^{-4}$\\
\hline
\multicolumn{6}{c}{$\tan  \beta = 20$} \\ \hline

$200$ &$M_{12}$  & $178$ & $3960 $& $15.1$ &$4.48\times 10^{-2}$\\
&$M_{ij}$  & $498 $& $14500$ & $22.2 $ &$3.43\times 10^{-2}$\\
\hline

$400$&$M_{12}$  & $104$& $6680$ & $6.94$ &$1.56\times 10^{-2}$\\
&$M_{ij}$  & $143  $& $11900$ & $7.14 $ &$1.20 \times 10^{-2} $\\
\hline

$800$&$M_{12}$  & $6.99$& $1400$ & $1.02$ &$5.00\times 10^{-3}$\\
&$M_{ij}$  & $7.96 $& $2380$ & $0.891 $ &$3.34\times 10^{-3}$\\
\hline

\hline
\multicolumn{6}{c}{$\tan  \beta = 50$} \\ \hline

$200$ &$M_{12}$  & $961$ & $3960 $& $75.0$ &$2.42\times 10^{-1}$\\
&$M_{ij}$  & $2770 $& $14500$ & $115 $ &$1.91\times 10^{-1}$\\
\hline

$400$&$M_{12}$  & $563$& $6680$ & $36.2$ &$8.43\times 10^{-2}$\\
&$M_{ij}$  & $ 792 $& $11900$ & $38.5 $ &$6.66\times 10^{-2}$\\
\hline

$800$&$M_{12}$  & $38.7$& $1400$ & $5.58$ &$2.76 \times 10^{-2}$\\
&$M_{ij}$  & $44.7 $& $2380$ & $4.96 $ &$ 1.87\times 10^{-2}$\\
\hline
\end{tabular}
\end{table}

\bigskip

Furthermore, we have studied the effects of SUSY particles on the
$\phi^0b\bar{b}$ Yukawa couplings at large $\tan \beta$.
The SUSY contributions can be described with an effective Lagrangian
and a function $\Delta_b$
\cite{Hall:1993gn,Carena:1994bv,Pierce:1996zz,Carena2006}
such that  the bottom quark mass in Yukawa couplings becomes 
\begin{eqnarray*}
m_b \rightarrow \frac{m_b}{1 + \Delta_b}
\end{eqnarray*}
where SUSY QCD corrections lead to 
\begin{eqnarray*}
\Delta_b = \Delta^{\tilde{b}}_b = 
\frac{2\alpha_s}{3\pi} m_{\tilde g} \mu \tan\beta
 I(m_{\tilde b_1}, m_{\tilde b_2}, m_{\tilde g}) \quad 
\end{eqnarray*}
for bottom squarks and gluinos, and the auxiliary function is
\begin{eqnarray*}
I(a,b,c) = -\frac{1}{(a^2-b^2)(b^2-c^2)(c^2-a^2)}(a^2 b^2\ln\frac{a^2}{b^2}+b^2 c^2\ln\frac{b^2}{c^2}+c^2 a^2\ln\frac{c^2}{a^2}) \quad .
\end{eqnarray*}
Then the cross section can be estimated with a simple formula \cite{Carena2006}
\begin{eqnarray*}
\sigma(pp \to b\phi^0 +X)\times B(\phi^0 \to b\bar{b})
 \simeq \sigma_{SM}(pp \to bH +X)
        \times \frac{\tan^2\beta}{(1+\Delta_b)^2}
        \times \frac{9}{(1+\Delta_b)^2 +9}.
\end{eqnarray*}
In our analysis of SUSY effects, we adopt the
conventions in Refs.~\cite{Dawson:2007ur,Dawson:2007wh} and 
have used a more complete estimate, including the effects
of the modified Higgs width in the full BWR calculation.
Table II shows the cross section including 
(a) no SUSY effects, 
(b) contributions from bottom squarks and gluinos, and
(c) contributions from bottom squarks and gluinos as well as 
from top squarks and Higgsinos.
The top squark/Higgsino loops give an additional
effective correction to $m_b$,
\begin{eqnarray*}
\Delta^{\tilde{t}}_b = 
\frac{\alpha_t}{4\pi} A_t \mu \tan\beta
 I(m_{\tilde t_1}, m_{\tilde t_2}, \mu) \quad ,
\end{eqnarray*}
where $\alpha_t \equiv \lambda_t^2/4\pi$ 
($\lambda_t = \sqrt{2}m_t/v_2$ being the top Yukawa coupling), and
$A_t$ is the trilinear Higgs-stop coupling.
It is clear that SUSY effects reduce the Higgs cross section 
for a positive $\mu$ while they enhance the Higgs cross section 
for a negative $\mu$. 
The effect of the Higgsino/stop loops is highly dependant
on the size of $A_t$. We present two scenarios, $M_h^{max}$ and no-mixing, as 
defined in Ref.~\cite{Carena2006}. In the former the Higgsino/stop contribution
is comparable to the gluino/bottom-squark term, in the latter it is almost 
negligible.


\begin{table}
\label{Delta-b}
\caption[]{
Effect of $\Delta_b$ in $M_h^{max}$ (no mixing) scenario. 
Cross sections in fb for
$pp \to b \phi^0 \to bb\bar{b} + X$ using high $p_T$ 
($ > 70$ GeV) cuts. Tagging efficiencies have not been applied.
}
\begin{tabular}{|c c|c |c |c |}
\hline
\bf{Mass}(GeV) & & $\mathbf{\Delta_b = 0}$ & $\mathbf{\tilde{g} / \tilde{b} }$& $\mathbf{\tilde{g} / \tilde{b} +\tilde{H} / \tilde{t}}$ \\
\hline
\multicolumn{5}{c}{$\tan  \beta = 10$} \\ \hline
 
$200$& $\mu = +200$ & $698(708)$ & $646(658)$ & $633(656)$ \\ 
& $\mu = -200$ & $699(703)$ & $745(755)$ & $761(753)$ \\\hline

$400$& $\mu = +200$ & $155(155)$ & $143(144)$ & $140(145)$ \\ 
& $\mu = -200$ & $156(155)$ & $168(167)$ & $172(168)$ \\\hline

$800$& $\mu = +200$ & $7.90(7.91)$ & $7.28(7.31)$ & $7.07(7.31)$ \\ 
& $\mu = -200$ & $7.87(7.93)$ & $8.63(8.56)$ & $8.86(8.60)$ \\ 

\hline
\multicolumn{5}{c}{$\tan  \beta = 50$} \\ \hline

$200$& $\mu = +200$ & $16400(16400)$ & $12200(12200)$ & $11000(12200)$ \\ 
& $\mu = -200$ & $16400(16300)$ & $22600(22600)$ & $25800(22600)$ \\\hline

$400$& $\mu = +200$ & $4120(4120)$ & $3060(3060)$ & $2750(3060)$ \\ 
& $\mu = -200$ & $4130(4120)$ & $ 5730(5730)$ & $6560(5720)$ \\\hline

$800$& $\mu = +200$ & $233(233)$ & $172(172)$ & $154(172)$ \\ 
& $\mu = -200$ & $ 233(233)$ & $325(325)$ & $373(325)$ \\
\hline

\end{tabular}
\end{table}

\section{The Discovery Potential at the Fermilab Tevatron}

To study the discovery potential of Higgs decays into bottom quark
pairs at the Fermilab Tevatron Run II, we require
\begin{itemize}
\item[(i)] three $b$ quarks or 3 jets (at least two $b$ jets) 
with $p_T > 15$ GeV or $p_T(j_1,j_2,j_3) > 50,30,15$ GeV, 
$|\eta(b,j)| < 2.0$, and a $b-$tagging efficiency 
$\eps_b = 50\%$~\cite{CDF-Abulencia:2005aj},
\item[(ii)] the angular separation between each pair of jets should be
$\Delta R > 0.4$ \cite{cdf2008},
\item[(iii)]
the invariant mass of the reconstructed bottom quark pairs should be within the
mass window of the Higgs mass with $\Delta M_{bb} = {\rm MAX}(0.1 \times M_\phi,  20 {\rm GeV})$.
\end{itemize}

Figure 5 show the $5\sigma$ discovery contours for the MSSM Higgs
bosons, where the discovery region is the part of the parameter space
above the curves. The discovery contours for $\Delta M_{bb} = 0.10
\times M_\phi$ \cite{tevforlhc} are comparable to those presented 
in this figure.

We find that the Tevatron Run II can discover neutral Higgs bosons 
in the MSSM for a value of $\tan\beta$ slightly less than 30 with an
integrated luminosity of 4 fb$^{-1}$ and $M_A < 120$ GeV. For
$\tan\beta \sim 50$, the Tevatron Run II will be able to discovery the
Higgs bosons up to $M_A \sim 160$ GeV with $L = 4$ fb$^{-1}$, and 
up to $M_A \sim 200$ GeV with $L = 20$ fb$^{-1}$. Our results are
consistent with those found in Refs.~\cite{Carena:1998gk,cdf2008,Draper:2009fh}.

 
\begin{figure}[htb]
\centering\leavevmode
\epsfxsize=3.6in
\epsfbox{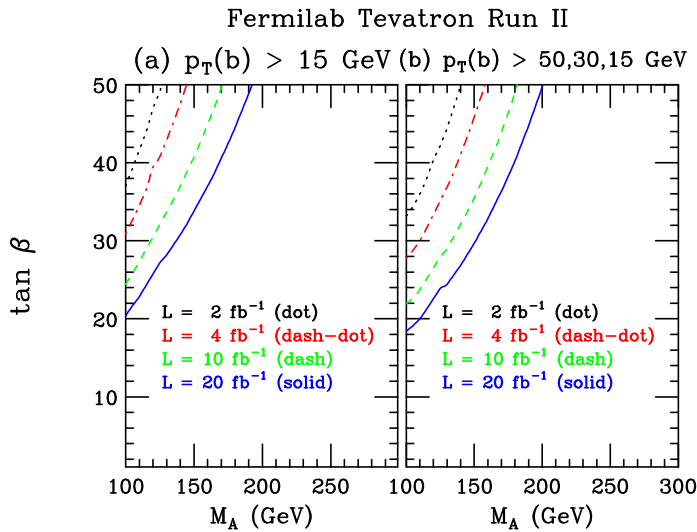}
\caption[]{
The $5\sigma$ discovery contours at the Fermilab Tevatron Run II
for an integrated luminosity ($L$) of 4 fb$^{-1}$, 10 fb$^{-1}$, 
20 fb$^{-1}$ in the $M_A$ versus $\tan\beta$ plane.  
The signal includes $\phi^0 = A^0$ and $h^0$ for $M_A < 125$ GeV, 
and $\phi^0 = A^0$ and $H^0$ for $M_A \ge 125$ GeV.
The discovery region is the 
part of the parameter space above the contours.}
\label{fig:hbbb-contour2}
\end{figure}

\section{Conclusions}

The associated production of a Higgs boson with a bottom quark,
followed by the Higgs decay into bottom quark pairs,
is a promising channel for the discovery of 
the neutral Higgs bosons in the minimal supersymmetric standard model 
at the LHC. The $A^0$ and the $H^0$ should be observable in a large region 
of parameter space with $\tan\beta \agt 10$.
The associated final state of $b\phi^0 \to b b\bar{b}$ 
could discover the $A^0$ and the $H^0$ at the LHC 
with an integrated luminosity of 30 fb$^{-1}$ if $M_A \alt 800$ GeV.
At a higher luminosity of 300 fb$^{-1}$, the discovery region in 
$M_A$ is expanded up to $M_A = 1$ TeV for $\tan\beta \sim 50$.

In our analysis, we apply a mass cut, requiring the reconstructed 
Higgs mass to lie in the mass window $M_\phi \pm \Delta M_{bb}$.
We note that improvements in the discovery potential will be possible 
by narrowing $\Delta M_{bb}$ if the bottom quark pair mass resolution 
can be improved. In regions of high mass and low $\tan \beta$ the ratio of 
signal to background events is very low. Discovery in these regions would 
require either excellent understanding of backgrounds in order to lower systematic
 errors below the few percent level, or better discrimination between signal and background due to narrower $\Delta M_{bb}$ or improved $b$-tagging. Our results using three $b$'s are more promising than
those found in previous studies based on $4b$ analyses 
\cite{Vega,Yuan,CMS,ATLAS-thesis}.

The discovery of the associated final state of 
$b\phi^0 \to b b\bar{b}$ along with 
$b\phi^0 \to b\tau^+\tau^-$ \cite{hbll} and 
$b\phi \to b\mu^+\mu^-$ \cite{hbmm}
will provide information about the Yukawa couplings of 
$f\bar{f}\phi^0; f = b, \tau, \mu$, for fermions with $t_3 = -1/2$. 
Furthermore, the muon pair
channel can also be observable in a significantly large region 
and the muon pair channel will provide a good opportunity to 
precisely reconstruct the masses for MSSM Higgs bosons 
\cite{hbbmm,hbmm,Nikita}.
In concert, this family of channels may provide an excellent window
on the Yukawa sector of the MSSM.

\section*{Acknowledgments}

We are grateful to Michelangelo Mangano for beneficial instruction and
discussions. 
C.K. thanks the Physics Division of CERN 
for hospitality and support during a sabbatical visit. 
This research was supported 
in part by the U.S. Department of Energy
under Grant No.~DE-FG02-04ER41305.
    
\newpage


\end{document}